\documentclass{IEEEtaes}

\usepackage{color,array,amsthm}
\usepackage{graphicx}

\usepackage{url}            
\usepackage{cite}
\usepackage{algorithm}
\usepackage{algorithmic}
\usepackage{amsmath}
\usepackage{mathrsfs}
\usepackage{amssymb}
\usepackage{caption}
\usepackage{subfig}
\usepackage{booktabs}       

\paper{1234567}

\begin{document}
\bibliographystyle{IEEEtaes}

\title{A Fast Algorithm for Onboard Atmospheric Powered Descent Guidance}

\author{Yushu Chen}
\affil{Department of Computer Science and Technology, 
	Tsinghua University, Beijing, China. Also at National Supercomputing Center in Wuxi, Jiangsu, China}
\author{Guangwen Yang}
\affil{Department of Computer Science and Technology, 
	Tsinghua University, Beijing, China. Also at National Supercomputing Center in Wuxi, Jiangsu, China and Zhejiang Lab, Hongzhou, China} 

\author{Lu Wang}
\author{Haipeng Chen}
\author{Qingzhong Gan}
\affil{Shanghai Aerospace Control Technology Institute, Shanghai, China}

\author{Quanyong Xu}
\affil{School of Aerospace Engineering, Tsinghua University, Beijing, China}


\receiveddate{The paper is accepted by IEEE Transactions on Aerospace and Electronic Systems, and the copyright has been transferred to IEEE. The formal paper can be accessed at \url{https://ieeexplore.ieee.org/document/10122793}. The citation is 'Y. Chen, G. Yang, L. Wang, H. Chen, Q. Gan and Q. Xu, "A Fast Algorithm for Onboard Atmospheric Powered Descent Guidance," in IEEE Transactions on Aerospace and Electronic Systems, doi: 10.1109/TAES.2023.3271961.' \\
This work was supported in part by the National Key Basic Research Program of China under Grant 2020YFB0204800 and the Key Research Project of Zhejiang Lab under Grant 2021PB0AC01.}

\corresp{{\itshape (Corresponding author: G. Yang)}.}

\authoraddress{Yushu Chen and Guangwen Yang are with the Department of Computer Science and Technology, 
	Tsinghua University, Beijing, 100084, China (e-mail:  \href{mailto:chenyushu@mail.tsinghua.edu.cn}{chenyushu@mail.tsinghua.edu.cn}, \href{mailto:ygw@mail.tsinghua.edu.cn}{ygw@mail.tsinghua.edu.cn}). They are also with National Supercomputing Center in Wuxi, Jiangsu, 214072, China. Guangwen Yang is also at Zhejiang Lab, Hongzhou, 311121, China. Lu Wang, Haipeng Chen, and Qingzhong Gan are with Shanghai Aerospace Control Technology Institute, Shanghai, 201108, China (e-mail: \href{mailto:wanglu1255@163.com}{wanglu1255@163.com}, \href{mailto:key\_chenhaipeng@sina.com}{key\_chenhaipeng@sina.com}, \href{mailto:hit2010gqz@163.com}{hit2010gqz@163.com}). Quanyong Xu is with School of Aerospace Engineering, Tsinghua University, Beijing, 100084, China (e-mail: \href{mailto:xuquanyong@tsinghua.edu.cn}{xuquanyong@tsinghua.edu.cn}).}


\markboth{YUSHU CHEN ET AL.}{FAST ALGORITHM FOR ATMOSPHERIC PDG}
\maketitle

\begin{abstract}Atmospheric powered descent guidance (APDG) can be solved by successive convexification; however, its onboard application is impeded by high computational cost. When aerodynamic forces are ignored, powered descent guidance (PDG) can be converted to a single convex problem. In contrast, APDG has to be converted into a sequence of convex subproblems, each of which is significantly more complicated. Consequently, the computation increases sharply. A fast real-time interior point method was presented to solve the correlated convex subproblems efficiently onboard in the work. The main contributions are as follows: Firstly, an algorithm was proposed to accelerate the solution of linear systems that cost most of the computation in each iterative step by exploiting the specific problem structure. Secondly, a warm-starting scheme was introduced to refine the initial value of a subproblem with a rough approximate solution of the former subproblem, which lessened the iterative steps required for each subproblem. The method proposed reduced the run time by a factor of 9 compared with the fastest publicly available solver tested in Monte Carlo simulations to evaluate the efficiency of solvers. Runtimes on the order of 0.6 s are achieved on a radiation-hardened flight processor, which demonstrated the potential of the real-time onboard application.
\end{abstract}


\section{INTRODUCTION}
P{\scshape owered} descent guidance (PDG) generates thrust magnitude and direction commands during the powered descent phase in the landing mission of a vertical takeoff and vertical landing (VTVL) reusable rocket. The vehicle uses the engine for retropropulsion following the commands, which realizes fuel-optimal, soft, vertical, and pinpoint landing and satisfies certain constraints. 

The PDG problem needs to be solved onboard in real-time. The computational efficiency is critical because the initial value has to be predicted for time offset to compensate for the run time, and random or unpredictable factors during time offset may lead to unacceptable errors in the real trajectory. However, typically, the radiation-hardened flight processors are significantly slower than contemporary general processors, which requires algorithms to be highly efficient. The challenge of the run time is more serious in landing missions on the earth or a planet with a thick atmosphere, where aerodynamic forces have a significant non-linear effect on the trajectory and further complicate the atmospheric powered descent guidance (henceforth referred to as APDG) problem.

The powered descent architecture developed for the Apollo Program exemplifies PDG strategies. Trajectories are represented as polynomials parameterized in time, and the coefficients are solved to meet specified terminal conditions in Refs. \cite{CHERRY1964,KLUMPP1974}. The method is adapted for Mars landing missions \cite{Wong2006,Singh2007,Sell2014}. Although polynomial guidance is highly efficient in computation, which enables landing missions with very limited computing power, it is not fuel-optimal \cite{Ross2004}. More complicated methods are applicable as processor performance increases, which arouses interest to generate fuel-optimal trajectories and includes more realistic constraints. More complicated methods are applicable as processor performance increases, which arouses interest to generate fuel-optimal trajectories and includes more realistic constraints.

Fuel-optimal PDG is the hotspot of academic research recently \cite{Song2020}. Lossless convexification transforms fuel-optimal PDG into a convex optimization problem, which allows for the global optimal descent trajectory to be computed with guaranteed convergence \cite{Acikmese2007, Blackmore2010, Accikmecse2013}. Solving the problem by convex optimization further facilitates enforcing convex path constraints such as minimum glide slope and maximum off-vertical thrust direction \cite{Acikmese2007, Blackmore2010, Accikmecse2013, Reynolds2019, Lee2017}. The guidance for fuel-optimal large diverts (G-FOLD) algorithm \cite{Scharf2017}, solving fix-time 3D fuel-optimal PDG through lossless convexification, has been demonstrated by test flights. The original problem is transformed into a convex problem in G-FOLD, or more specifically, a second-order cone-programming (SOCP) problem \cite{Boyd2004}. A customized real-time interior point solver (Bsocp) \cite{Dueri2017} is developed to solve SOCPs onboard.

Many promising approaches are developed for fuel-optimal PDG besides lossless convexification. For example, an indirect method \cite{Lu2018, Lu2018adaptive, Lu2019, Lu2020, Johnson2019} is used to obtain the optimal descent trajectory and the optimal burn time by solving a multivariate root-finding problem, which can compute onboard efficiently. It also accommodates different problem formulations and terminal constraints. On the other hand, the indirect method is difficult to enforce inequality constraints such as minimum glide slope and thrust pointing direction and does not enjoy theoretically guaranteed convergence. A stochastic process method \cite{Ridderhof2021} is proposed to combine open-loop PDG solutions with closed-loop control, which adjusts bounds on the feed-forward optimal thrust magnitude command to allow for sufficient feedback authority. Machine learning-based approaches \cite{Sanchez2018, Gaudet2020, You2020, Hendrix2022} are efficient in computation and emerging in this area. 

Nonlinear aerodynamic forces complicate dynamics for reusable-rocket landing on earth or a planet with a thick atmosphere, which makes the APDG problem more difficult to solve. A successive convexification approach \cite{Szmuk2016} is proposed in the presence of aerodynamic drag and new types of non-convex control constraints. The original non-convex problem is transformed into a sequence of SOCPs by lossless convexification and successive linearization \cite{Liu2016,Miao2021}. The approach is extended to solve more complex problems. Aerodynamic lift and drag are considered in the successive solution procedure in a 2D problem \cite{Liu2019}. Ref. \cite{Gan2018} considers the variation of atmospheric density with altitude. Altitude is used as an independent variable instead of time \cite{Yang2020} to deal with free final time and incorporates altitude-dependent glide-slope and thrust-direction constraints. Ref. \cite{Szmuk2020} solves a generalized six-degree-of-freedom powered descent guidance problem with aerodynamic lift and drag, which also solves the engine ignition time and final time and tackles enforced constraints conditionally. Convex and non-convex contributions are processed separately to maximize computational efficiency \cite{Sagliano2021}. The computation complexity of these approaches scale poorly with the presence of nonlinear aerodynamic forces. This results in unrealistic computation requirements for many non-trivial landing scenarios. 

SOCPs are typically solved by interior point method (IPM) solvers, e.g., SeDuMi \cite{Sturm1999}, SDPT3 \cite{Toh1999}, MOSEK \cite{Andersen2000}, ECOS \cite{Domahidi2013}, and Coneprog \cite{Coneprog}. Although no one solver performs better than the others on every problem \cite{Grant2014}, MOSEK and ECOS are competitive for fuel-optimal PDG \cite{Dueri2017, Yang2020}. A customized IPM solver (Bsocp) is developed to improve fuel-optimal PDG without aerodynamic forces \cite{Dueri2017}. Bsocp is more than twice faster than ECOS in small problems, but ECOS is better for large problems. The density of the equity-constraint coefficient matrix (hereinafter called equity constraint density) also significantly affects the computational efficiency besides the number of solution variables. When aerodynamic forces are introduced, the equity constraint density increases. It leads to substantial growth of the density of the coefficient matrix in linear systems to the solver for each iterative step in the algorithm \cite{Andersen2003, Wang2003} of Bsocp, which severely deteriorates the computational efficiency. 

Although ECOS performs better as the equity constraint density increases, a more efficient solver for SOCPs is still needed urgently. On the other hand, the sequence of SOCPs in successive convexification, each of which is called a subproblem, is closely related. Consequently, warm-starting, not supported in MOSEK and ECOS, has the potential to gain significant acceleration by exploiting the correlation between subproblems. However, it is widely perceived that warm-starting is hard for IPMs \cite{Skajaa2013,Potra2000} because IPMs typically converge slowly and suffer from rapid variation of the gradients and Hessian matrice of the barrier functions when the solutions are close to the boundary of the inequity constraints and not well-centered.

A fast interior point algorithm was proposed for correlated SOCP subproblems in the successive convexification of APDG in the work. The method, an improved version of a classic IPM \cite{Andersen2003, Wang2003}, solves the homogeneous self-dual embedding problem \cite{Nemirovskii1993} with Nesterov-Todd scaling \cite{Nesterov1997} and Mehrotra’s predictor-corrector technique \cite{Mehrotra1991, Mehrotra1992}. The main contributions are as follows: 1) Linear systems were reformulated to exploit the sparse structure of the specific problem, which significantly accelerated the solution of linear systems occupying the major computation in each iteration. 2) A warm-starting scheme was introduced for acceleration using the correlation between subproblems, which enabled each subproblem to be run for only a few iterations. The solver developed (called FSOCP) was applied to solve APDG using the successive convexification approach \cite{Szmuk2016}, where nonlinear aerodynamic drag is considered. The average runtime of FSOCP was 10.5\% of MOSEK and 3.4\% of ECOS in the Monte Carlo simulation to evaluate the efficiency of solvers. Runtimes on the order of 0.6 s are achieved on a radiation-hardened P2020NXE2KHC flight processor, showing that the method is onboard applicable to solve fuel-optimal APDG.

The work is organized as follows: Section \uppercase\expandafter{\romannumeral2} introduces a classic IPM for SOCPs, which can be applied to solve convex subproblems obtained in successive convexification of APDG and serves as the framework of the algorithm proposed. An approach is presented in Section \uppercase\expandafter{\romannumeral3} to accelerate the solution of linear systems by exploiting the sparse structure of the specific problem. Section \uppercase\expandafter{\romannumeral4} proposes the warm-starting scheme to utilize the correlation between subproblems. Section \uppercase\expandafter{\romannumeral5} compares numerical results of different SOCP solvers in APDG, and Section \uppercase\expandafter{\romannumeral6} summarizes the work.

\section{A Brief Review of a Classic IPM for SOCPs}

APDG is transformed into a sequence of SOCPs in the successive convexification approach \cite{Szmuk2016}. This section introduces a classic interior point method to solve SOCPs, which serves as the framework of the algorithm proposed.

\subsection{SOCP Problem}

SOCP is a convex optimization problem that minimizes a linear function over the intersection of an affine set and the Cartesian product of linear cones (LCs) and second-order (Lorentz) cones (SOCs). SOCP includes linear programming (LP), convex quadratic programming (QP), and quadratically constrained convex quadratic programming (QCQP) as special cases, but are less general than semidefinite programming (SDP) \cite{Alizadeh2003}. It can be applied in engineering and quantitative finance, e.g., filter design, antenna-array weight design, truss design, and portfolio optimization \cite{Miguel1998}.

Linear cones and second-order cones are defined as follows.

\textbf{Definition 1}: A linear cone is a convex set defined by
\begin{equation}
	\mathcal{K}_{L}^{n}=\{\mathbf{v }\in {{\mathbf{R}}^{n}}:\mathbf{v }\ge 0\}, n\ge 1, \\
	\label{LC}
\end{equation}
where $n$ is the number of dimensions; $\ge$ an elementwise operator.

\textbf{Definition 2}: A second-order (Lorentz) cone is a convex set defined by
\begin{equation}
	\mathcal{K}_{S}^{n}=\{\mathbf{v }\in {{\mathbf{R}}^{n}}:{{\mathbf{v }}_{1}}\ge \left\| {{\mathbf{v }}_{2:n}} \right\|\}, n\ge2,
	\label{SOC}
\end{equation}
where $\|\cdots\|$ is used instead of $\|\cdots\|_2$ for simplicity. The 1-dimensional SOC is defined as $\mathcal{K}_{S}^{1}=\mathcal{K}_{L}^{1}$. The superscript of the dimension number is omitted if it is not required to be specified in a linear cone or a SOC.

The standard form of SOCP is defined as

\textbf{Problem 1}: The standard form of SOCP problem
\begin{equation}
	\begin{aligned}
		&\underset{\mathbf{x}}{\text{minimize}}~\mathbf{c}^T\mathbf{x},\\
		&\text{s.t.}\\
		&\mathbf{A}\mathbf{x}=\mathbf{b},\\
		&\mathbf{x}\in\mathcal{K},\\
		&\mathcal{K} =\mathcal{K}_{L}^{l}\times \mathcal{K}_{S}^{{{n}_{1}}}\times \mathcal{K}_{S}^{{{n}_{2}}}\cdots \times \mathcal{K}_{S}^{{{n}_{m}}}.
		\label{SOCP}
	\end{aligned}
\end{equation}
$\mathbf{A} \in \mathcal{R}^{p\times n}$ is the equity-constraint coefficient matrix with $\text{rank}(\mathbf{A})=p\le n$. $\mathbf{x} \in \mathcal{R}^n$ is the solution variable, and $\mathbf{c}\in \mathcal{R}^n$ is the coefficient vector in the objective function. Each element of $\mathbf{x}$ is constrained in a LC or a SOC. The LC is assumed to be arranged at first in the solution variable without loss of generality, followed by the SOCs. $\mathcal{L}^l_L$ is viewed as $l$ linear cones of dimension 1 hereinafter for convience.

SOCP subproblems generated by the successive convexification of APDG can be transformed to the standard form by the canonicalization method described in Ref. \cite{Dueri2017}.

\subsection{Classic IPM for SOCPs}

The classic method introduced \cite{Andersen2003, Wang2003} is a primal-dual path-following IPM solving the primal and dual problems of SOCP simultaneously. It has a rich verification history and forms the basis of a real-time custom IPM solver without aerodynamic forces that has been flight-tested \cite{Dueri2017}. The key idea of the algorithm is tracking the central path loosely to the optimal solution of a bilinear homogeneous self-dual (HSD) problem \cite{Nemirovskii1993}. The HSD problem fully describes the properties of the primal problem and the dual problem and can be initialized trivially. The HSD model is optimized by a Newton method, and Nesterov-Todd scaling \cite{Nesterov1997} is applied to make the problem numerically well-conditioned. Besides, Mehrotra’s predictor-corrector method \cite{Mehrotra1991, Mehrotra1992} is used to correct the search directions. We merely describe the algorithm without deviation because of limited pages. 

The primal problem (Problem 1) and its dual problem are closely related, so modern IPMs use the information from one to progress the other by solving them simultaneously \cite{Skajaa2013}. The dual problem of SOCP is given by

\textbf{Problem 2}: The dual problem of SOCP
\begin{equation}
	\begin{aligned}
		&\underset{\mathbf{y}}{\text{maximize}}~\mathbf{b}^T\mathbf{y}\\
		&\text{s.t.}\\
		&\mathbf{A}^T\mathbf{y}+\mathbf{s}=\mathbf{c},\\
		&\mathbf{s}\in\mathcal{K}.\\
		\label{SOCP_DUAL}
	\end{aligned}
\end{equation}
$\mathbf{y} \in \mathcal{R}^p$  and $\mathbf{s} \in \mathcal{R}^n$ are the solution variables of the dual problem, and $\mathbf{b} \in \mathcal{R}^p$ is the coefficient vector in the dual objective function.

In the algorithm, the approach introduced by Nesterov and Todd \cite{Nesterov1997} is applied to scale the searching directions, which improves numerical stability and accelerates convergence with low computational cost. The scaling variables for the i$^{th}$ cone include positive scalar $\theta$ and symmetric matrix $\mathbf{G}$.
\begin{equation}
	{{\left( {{\theta }^{(i)}} \right)}^{2}} \triangleq 
	\sqrt{\frac{{{\left( {{\mathbf{s}}^{(i)}} \right)}^{T}}{{\mathbf{Q}}^{(i)}}{{\mathbf{s}}^{(i)}}}{{{\left( {{\mathbf{x}}^{(i)}} \right)}^{T}}{{\mathbf{Q}}^{(i)}}{{\mathbf{x}}^{(i)}}}},
\end{equation}
where $\mathbf{Q}^{(i)} \triangleq 1$ for LCs and  $\mathbf{Q}^{(i)} \triangleq \text{diag}(1,-1,\cdots,-1)$ for SOCs. $\mathbf{G}^{(i)}$  is also 1 for LCs and
\begin{equation}
	{{\mathbf{G}}^{(i)}}\triangleq -{{\mathbf{Q}}^{(i)}}+\frac{({{\mathbf{e}}^{(i)}}+{{\mathbf{q}}^{(i)}}){{(\mathbf{e}^{(i)}+{{\mathbf{q}}^{(i)}})}^{T}}}{1+{{\left( {{\mathbf{e}}^{(i)}} \right)}^{T}}{{\mathbf{q}}^{(i)}}}
	\label{DEF_G}
\end{equation}
for SOCs, where
\begin{equation}
	{{\mathbf{q}}^{(i)}}\triangleq \frac{{{\mathbf{s}}^{(i)}}/{{\theta }^{(i)}}+{{\theta }^{(i)}}{{\mathbf{Q}}^{(i)}}{{\mathbf{x}}^{(i)}}}{\sqrt{2\left( {{\mathbf{x}}^{(i)T}}{{\mathbf{s}}^{(i)}}+\sqrt{{{\mathbf{x}}^{(i)T}}{{\mathbf{Q}}^{(i)}}{{\mathbf{x}}^{(i)}}{{\mathbf{s}}^{(i)T}}{{\mathbf{Q}}^{(i)}}{{\mathbf{s}}^{(i)}}} \right)}},
\end{equation}
and $\mathbf{e}^{(i)}=(1,0,\cdots, 0)^T$.
For convience, we also define the unit vector as
\begin{equation}
	\mathbf{e}\triangleq {{\left( {{\left( {{\mathbf{e}}^{(1)}} \right)}^{T}},\cdots ,{{\left( {{\mathbf{e}}^{(l+m)}} \right)}^{T}} \right)}^{T}}.
\end{equation}

Then, the cone is scaled as 
\begin{equation}
	{{\mathbf{\bar{x}}}^{(i)}} \triangleq {{\theta }^{(i)}}{{\mathbf{G}}^{(i)}}{{\mathbf{x}}^{(i)}}, ~ 
	{{\mathbf{\bar{s}}}^{(i)}} \triangleq {{({{\theta }^{(i)}}{{\mathbf{G}}^{(i)}})}^{-1}}{{\mathbf{s}}^{(i)}}.
\end{equation}
Solution variables $\mathbf{x}$ and $\mathbf{s}$ are scaled as
\begin{equation}
	\mathbf{\bar{x}} \triangleq {{\mathbf{D}}^{-1}}\mathbf{x},~
	\mathbf{\bar{s}} \triangleq \mathbf{Ds},
	\label{DEF_NT}
\end{equation}
where
\begin{equation}
	\begin{aligned}
		&\mathbf{D}\triangleq {{(\Theta \mathbf{G})}^{-1}},\\
		&\Theta \triangleq \text{blkdiag}\left( {{\theta }^{(1)}},\cdots ,{{\theta }^{(l)}},{{\theta }^{(l+1)}}{{\mathbf{I}}_{{{n}_{1}}}},\cdots,  {{\theta }^{(l+m)}}{{\mathbf{I}}_{{{n}_{m}}}} \right),\\
		&\mathbf{G}\triangleq \text{blkdiag}\left( {{\mathbf{G}}^{(1)}},\cdots ,{{\mathbf{G}}^{(l+m)}} \right).
	\end{aligned}
	\label{DEF_D}
\end{equation}

Most computation in the classic IPM is used to solve linear systems, which obtains the Newton direction. Denote the current estimation of the solution by $\mathbf{z}=\left( \mathbf{x},\mathbf{y},\mathbf{s},\kappa ,\tau  \right)$, where $\kappa,\tau \in \mathcal{K}$ are relax variables. The linear system is
\begin{equation}
	\begin{aligned}
		& \mathbf{A}\Delta \mathbf{x}-\mathbf{b}\Delta \tau =\ {{\mathbf{w}}_{1}} \\ 
		& -{{\mathbf{A}}^{T}}\Delta \mathbf{y}+\mathbf{c}\tau -\Delta \mathbf{s}={{\mathbf{w}}_{2}} \\ 
		& {{\mathbf{b}}^{T}}\Delta \mathbf{y}-{{\mathbf{c}}^{T}}\Delta \mathbf{x}-\Delta \kappa ={{w}_{3}} \\ 
		& \mathbf{\bar{X}D}\Delta \mathbf{s}+\mathbf{\bar{S}}{{\mathbf{D}}^{-1}}\Delta \mathbf{x}={{\mathbf{w}}_{4}} \\ 
		& \kappa \Delta \tau +\tau \Delta \kappa ={{w}_{5}},
		\label{LINSYS_SCAL}
	\end{aligned}
\end{equation}
where
\begin{equation}
	\begin{aligned}
		& {{\mathbf{w}}_{1}}\triangleq -\left( 1-\nu  \right)\left( \mathbf{Ax}-\mathbf{b}\tau  \right) \\ 
		& {{\mathbf{w}}_{2}}\triangleq -\left( 1-\nu  \right)\left( -{{\mathbf{A}}^{T}}\mathbf{y}+\mathbf{c}\tau -\mathbf{s} \right) \\ 
		& {{w}_{3}}\triangleq -\left( 1-\nu  \right)\left( {{\mathbf{b}}^{T}}\mathbf{y}-{{\mathbf{c}}^{T}}\mathbf{x}-\kappa  \right) \\ 
		& {{\mathbf{w}}_{4}}\triangleq \mu \nu \mathbf{e}-\mathbf{\bar{X}\bar{S}e}-{{\mathbf{E}}_{xs}} \\ 
		& {{w}_{5}}\triangleq \mu \nu -\kappa \tau -{{E}_{\kappa \tau }} \\ 
		& \mathbf{\bar{X}}\triangleq \text{mat}\left( {\mathbf{\bar{x}}} \right),\mathbf{\bar{S}}\triangleq \text{mat}\left( {\mathbf{\bar{s}}} \right) \\ 
		& \mu \triangleq \frac{\mathbf{x}_{{}}^{T}\mathbf{s}+\tau \kappa }{m+l+1}, 
	\end{aligned}
	\label{LINSYS_SCAL_TERMS}
\end{equation}
$\mathbf{E}_{xs}\in \mathcal{R}^n$ approximates second-order term $\Delta \mathbf{\bar{X}}\Delta \mathbf{\bar{S}e}$.

In equation (\ref{LINSYS_SCAL_TERMS}), $\text{mat}(\mathbf{h}), \mathbf{h} \in \mathcal{K}$ is the block arrowhead matrix associated with the cone constraint $\mathcal{K}$, which characterizes the complementarity condition in SOCP. It is defined as
\begin{equation}
	\begin{aligned}
		&\text{mat}(\mathbf{h}) \triangleq \text{blkdiag}\left(\text{arrow}\left(\mathbf{h}^{(1)}\right), \cdots, \text{arrow}\left(\mathbf{h}^{(l+m)}\right)\right), \\
		&\text{arrow}\left(\mathbf{h}^{(i)}\right)\triangleq 
		\left\{
			\begin{aligned}
			&{\mathbf{h}}^{(i)},\quad i \in \left\{1,\cdots, l\right\},\\
			&\begin{pmatrix}
				{\mathbf{h}_{1}^{(i)}} & \left(\mathbf{h}_{2:n_{i-l}}^{(i)}\right)^{T}  \\
				\mathbf{h}_{2:n_{i-l}}^{(i)} & {\mathbf{h}_1^{(i)}}{{\mathbf{I}}_{n_{i-l}-1}} 
			\end{pmatrix},\quad i \in \left\{l+1,\cdots, l+m\right\},
			\end{aligned}
		\right.
	\label{BLK_ARROW_MAT}
	\end{aligned}
\end{equation}
where $\mathbf{h} \in \mathcal{K}$ is a solution variable;  $\mathbf{h}^{(i)}$ a subvector of $\mathbf{h}$ in the linear cone if $i=0$  or the $i^{th}$ SOC if  $i\in \{1,\cdots, m\}$; $\text{blkdiag}(\cdots)$ a block diagonal matrix with given diagonal blocks; $\mathbf{I}$ the identity matrix.

\begin{algorithm}[tb]
	\caption{Overview of the classic IPM for SOCP}
	\label{ALG_CLASSIC}
	\textbf{Input}: problem parameters $\mathbf{A}$, $\mathbf{b}$, $\mathbf{c}$, and $\mathcal{K}$, initial value ${{\mathbf{z}}_{0}}=\left( {{\mathbf{x}}_{0}},{{\mathbf{y}}_{0}},{{\mathbf{s}}_{0}},{{\kappa }_{0}},{{\tau }_{0}} \right)$, maximum iteration number $N_{iter}$, and coefficient $\delta_0, \delta_1 \in (0,1)$,  ($\delta_0=0.995, \delta_1=0.9$ by default).\\
	\textbf{Output}: Solution $\mathbf{z}=(\mathbf{x},\mathbf{y},\mathbf{s},\kappa,\tau)$ and solver status.
	\begin{algorithmic}[1] 
		\STATE Initialize the current solution $\mathbf{z} = \mathbf{z}_{0}$;
		\FOR{$i=0,\cdots ,{{N}_{iter}}-1$}
		\STATE Set $\mathbf{E}_{xs}=0, {E}_{\kappa \tau }=0, \nu =0$;\\
		\COMMENT{Predictor}
		\STATE Solve linear system (\ref{LINSYS_SCAL}) to obtain the direction denoted as $\left(\Delta {{\mathbf{x}}_{p}},\Delta {{\mathbf{y}}_{p}},\Delta {{\mathbf{s}}_{p}},\Delta {{\kappa }_{p}},\Delta {{\tau }_{p}}\right)$;
		\STATE Calculate Maximum Newton step size $\alpha_p$;
		\STATE Set $\alpha_p=\min(\alpha_p,\delta_0)$;
		\STATE Set $\mathbf{E}_{xs}=\text{mat}(\Delta {{\mathbf{x}}_{p}})\text{mat}(\Delta {{\mathbf{x}}_{p}})\mathbf{e}$, 
		${E}_{\kappa \tau }=\Delta \kappa_{p} \Delta \tau_{p}$, 
		$\nu =\min \left( {{\delta }_{1}},{{\left( 1-{{\alpha }_{p}} \right)}^{2}} \right)\left( 1-{{\alpha }_{p}} \right)$;\\
		\COMMENT{Corrector}
		\STATE Solve linear system (\ref{LINSYS_SCAL}) to obtain the direction denoted as $\left(\Delta {{\mathbf{x}}_{c}},\Delta {{\mathbf{y}}_{c}},\Delta {{\mathbf{s}}_{c}},\Delta {{\kappa }_{c}},\Delta {{\tau }_{c}}\right)$;
		\STATE Calculate Maximum Newton step size $\alpha_c$;
		\STATE Set $\alpha_c=\min(\alpha_c,\delta_0)$;
		\STATE Set $\mathbf{z}=\mathbf{z}+\alpha_c \left(\Delta {{\mathbf{x}}_{c}},\Delta {{\mathbf{y}}_{c}},\Delta {{\mathbf{s}}_{c}},\Delta {{\kappa }_{c}},\Delta {{\tau }_{c}}\right)$;
		\STATE Exit and output the solution and solver status if stop criteria are satisfied;
		\ENDFOR
	\end{algorithmic}
\end{algorithm}

The classic IPM applies Mehrotra’s predictor-corrector method, which increases the efficiency by using a second-order correction of the search direction. The method is summarized as Algorithm \ref{ALG_CLASSIC}, where Maximum Newton step size is the maximum step size that keeps the updated solution in the cone constraint. Typically, cold-starting is performed using
\begin{equation}
	{{\mathbf{z}}_{0}}={{\mathbf{z}}^{c}}\triangleq \left( {{\mathbf{x}}^{c}},{{\mathbf{y}}^{c}},{{\mathbf{s}}^{c}},{{\kappa }^{c}},{{\tau }^{c}} \right)=\left( \mathbf{e},0,\mathbf{e},1,1 \right).
\end{equation}
See Ref. \cite{Andersen2003} for the details of the stop criteria.

\section{Accelerating the Solution of Linear Systems}

Solving linear system (\ref{LINSYS_SCAL}) consumes most of the computation in the classic IPM, which requires acceleration. The section presents an approach to reduce computation by reformulating linear systems, which exploits the sparse structure of the specific problem.

The linear system is transformed into two linear systems with the coefficient matrix $\mathbf{A}\mathbf{D}^2\mathbf{A}^T$  in the original algorithm \cite{Andersen2003, Wang2003, Dueri2017}, which symmetrizes the coefficient matrix and reduces dimensions. $\mathbf{A}$ is sparse when dynamics are relatively simple, and $\mathbf{D}$ is also sparse when SOCs are relatively few. Therefore, the method is highly efficient in applications such as PDG without aerodynamic forces. However, although $\mathbf{A}$ and $\mathbf{D}$ still have good sparsity structures in general when they are complicated by introducing aerodynamic forces, the number of nonzero elements (hereinafter called nnz) of $\mathbf{A}\mathbf{D}^2\mathbf{A}^T$ increases sharply so that the computational efficiency decreases seriously.

Our research is motivated by the following ideas: 1) $\text{nnz}(\mathbf{A}\mathbf{D}^2\mathbf{A}^T)$  is typically much larger than $2\text{nnz}(\mathbf{A})+\text{nnz}(\mathbf{D}^2)$  when $\mathbf{A}$ and $\mathbf{D}$ are sparse and $\mathbf{D}$ is block diagonal. 2) When overall nnz of the coefficient matrix is constant, the sparsity increases with increased dimensions, and the computation of LDL decomposition typically decreases. Consequently, $\mathbf{A}$, $\mathbf{D}^2$, and $\mathbf{A}^T$ are used as separate blocks in the coefficient matrix based on the two points above with higher dimensions and typically smaller nnz. Therefore, less computation is required compared to the methods using $\mathbf{A}\mathbf{D}^2\mathbf{A}^T$ as coefficient matrice in most cases.

Linear system (\ref{LINSYS_SCAL}) can be written in a matrix form as
\begin{equation}
	\left( \begin{matrix}
		\mathbf{A} & 0 & 0 & 0 & -\mathbf{b}  \\
		0 & -{{\mathbf{A}}^{T}} & -I & 0 & \mathbf{c}  \\
		-{{\mathbf{c}}^{T}} & {{\mathbf{b}}^{T}} & 0 & -1 & 0  \\
		\mathbf{\bar{S}}{{\mathbf{D}}^{-1}} & 0 & \mathbf{\bar{X}D} & 0 & 0  \\
		0 & 0 & 0 & \tau  & \kappa   \\
	\end{matrix} \right)\underbrace{\left( \begin{matrix}
			\Delta \mathbf{x}  \\
			\Delta \mathbf{y}  \\
			\Delta \mathbf{s}  \\
			\Delta \kappa   \\
			\Delta \tau   \\
		\end{matrix} \right)}_{\mathbf{u}}=\left( \begin{matrix}
		{{\mathbf{w}}_{1}}  \\
		{{\mathbf{w}}_{2}}  \\
		{{w}_{3}}  \\
		{{\mathbf{w}}_{4}}  \\
		{{w}_{5}}  \\
	\end{matrix} \right),
	\label{LINSYS_MAT0}
\end{equation}
where $\mathbf{u}$ is the increment of the solution.

Some important properties of Nesterov-Todd scaling are useful to simplify the linear system, which are introduced by Theorem 1.1.5 in Ref. \cite{Wang2003} as
\begin{equation}
	\begin{aligned}
		& \mathbf{s}={{\Theta }^{2}}{{\mathbf{G}}^{2}}\mathbf{x} \\ 
		& {{\mathbf{G}}^{-2}}=-\mathbf{Q}+2\left( \mathbf{Qq} \right){{\left( \mathbf{Qq} \right)}^{T}}.
	\end{aligned}
	\label{PROP_NT}
\end{equation}

Equation (\ref{PROP_NT}) is combined with (\ref{DEF_NT}), (\ref{DEF_D}), and (\ref{LINSYS_SCAL_TERMS}) to obtain
\begin{equation}
	\bar{\mathbf{X}}=\bar{\mathbf{S}}.
\end{equation}

Then, linear system (\ref{LINSYS_MAT0}) is equivalent to
\begin{equation}
	\underbrace{\left( \begin{matrix}
			0 & {{\mathbf{A}}^{T}} & \mathbf{I} & 0 & -\mathbf{c}  \\
			\mathbf{A} & 0 & 0 & 0 & -\mathbf{b}  \\
			\mathbf{I} & 0 & {{\mathbf{D}}^{2}} & 0 & 0  \\
			0 & 0 & 0 & 1 & \kappa /\tau   \\
			-{{\mathbf{c}}^{T}} & {{\mathbf{b}}^{T}} & 0 & -1 & 0  \\
		\end{matrix} \right)}_{{{\mathbf{B}}_{0}}}\underbrace{\left( \begin{matrix}
			\Delta \mathbf{x}  \\
			\Delta \mathbf{y}  \\
			\Delta \mathbf{s}  \\
			\Delta \kappa   \\
			\Delta \tau   \\
		\end{matrix} \right)}_{\mathbf{u}}=\underbrace{\left( \begin{matrix}
			-{{\mathbf{w}}_{2}}  \\
			{{\mathbf{w}}_{1}}  \\
			{{{\mathbf{\hat{w}}}}_{4}}  \\
			{{w}_{5}}/\tau   \\
			{{w}_{3}}  \\
		\end{matrix} \right)}_{{{\mathbf{w}}_{0}}},
	\label{LINSYS_MAT1}
\end{equation}
where ${{\mathbf{B}}_{0}}$, $\mathbf{u}$, and ${{\mathbf{w}}_{0}}$ are notations of the corresponding terms, and
\begin{equation}
    {{\mathbf{\hat{w}}}_{4}}=\mathbf{D}{{\left( {\mathbf{\bar{X}}} \right)}^{-1}}{{\mathbf{w}}_{4}}.
    \label{W_HAT_4}
\end{equation}
It is shown in \cite{Andersen2003} that any matrix-vector product involving the matrices $D$, $D^{-1}$, $\text{mat}(\cdot)$, and $\text{mat}(\cdot)^{-1}$ can be carried out in $O(n)$ complexity, where $n$ is the dimension of the matrix. Consequently, (\ref{W_HAT_4}) can be computed efficiently.

Since it is typically more convenient to handle a symmetric coefficient matrix than a non-symmetric one, matrix $\mathbf{B}_0$is transformed into
\begin{equation}
	\begin{aligned}
		&{{\mathbf{B}}_{0}}=\\
		&\underbrace{\left( \begin{matrix}
				0 & {{\mathbf{A}}^{T}} & \mathbf{I} & 0 & 0  \\
				\mathbf{A} & 0 & 0 & 0 & 0  \\
				\mathbf{I} & 0 & {{\mathbf{D}}^{2}} & 0 & 0  \\
				0 & 0 & 0 & 1 & 0  \\
				0 & 0 & 0 & 0 & 1  \\
			\end{matrix} \right)}_{\widehat{\mathbf{B}}}+\underbrace{\left( \begin{matrix}
				-\mathbf{c}  \\
				-\mathbf{b}  \\
				0  \\
				\kappa /\tau   \\
				-1/2  \\
			\end{matrix}\ \ \begin{matrix}
				0  \\
				0  \\
				0  \\
				0  \\
				1  \\
			\end{matrix} \right)}_{{{\mathbf{R}}_{1}}}{{\underbrace{\left( \begin{matrix}
						0  \\
						0  \\
						0  \\
						0  \\
						1  \\
					\end{matrix}\ \ \ \begin{matrix}
						-\mathbf{c}  \\
						\mathbf{b}  \\
						0  \\
						-1  \\
						-1/2  \\
					\end{matrix} \right)}_{{{\mathbf{R}}_{2}}}}^{T}},
	\end{aligned}
 \label{LINSYS_MAT1_REFORM}
\end{equation}
where $\widehat{\mathbf{B}}$, ${{\mathbf{R}}_{1}}$, and ${{\mathbf{R}}_{2}}$ are notations of the corresponding terms.

Then, linear system (\ref{LINSYS_MAT1}) is rewritten as
\begin{equation}
	\left( \widehat{\mathbf{B}}+{{\mathbf{R}}_{1}}\mathbf{R}_{2}^{T} \right)\mathbf{u}={{\mathbf{w}}_{0}}.
	\label{LINSYS_MAT2}
\end{equation}

The Sherman-Morrison formula is used to solve linear system (\ref{LINSYS_MAT2}), which obtains
\begin{equation}
	\mathbf{u}={{\widehat{\mathbf{B}}}^{-1}}{{\mathbf{w}}_{0}}-{{\widehat{\mathbf{B}}}^{-1}}{{\mathbf{R}}_{1}}{{\left( \mathbf{I}+\mathbf{R}_{2}^{T}{{\widehat{\mathbf{B}}}^{-1}}{{\mathbf{R}}_{1}} \right)}^{-1}}\mathbf{R}_{2}^{T}{{\widehat{\mathbf{B}}}^{-1}}{{\mathbf{w}}_{0}}.
	\label{LINSYS_Sherman_Morrison}
\end{equation}
Three linear systems need to be solved to compute increment $\mathbf{u}$ by (\ref{LINSYS_Sherman_Morrison}), and they are denoted as
\begin{equation}
	\begin{aligned}
		&\widehat{\mathbf{B}}{{\mathbf{u}}_{0}}={{\mathbf{w}}_{0}}\\				
		&\widehat{\mathbf{B}}{{\mathbf{u}}_{1}}={{\left( {{\mathbf{R}}_{1}} \right)}_{1}}\\
		&\widehat{\mathbf{B}}{{\mathbf{u}}_{2}}={{\left( {{\mathbf{R}}_{1}} \right)}_{2}},
	\end{aligned}
	\label{LINSYS_SYM}
\end{equation}
where vectors ${{\left( {{\mathbf{R}}_{1}} \right)}_{1}}$ and ${{\left( {{\mathbf{R}}_{1}} \right)}_{2}}$ are the first and second columns of ${{\mathbf{R}}_{1}}$, respectively; ${{\mathbf{u}}_{0}}$, ${{\mathbf{u}}_{1}}$, and ${{\mathbf{u}}_{2}}$ are intermediate variables. The structures of $\widehat{\mathbf{B}}$ and ${{\left( {{\mathbf{R}}_{1}} \right)}_{2}}$ are used to obtain
\begin{equation}
	{{\mathbf{u}}_{2}}={{\left( 0,0,\cdots ,0,1 \right)}^{T}}.
\end{equation}
Then, $\mathbf{u}$ can be computed as a linear combination of ${{\mathbf{u}}_{0}}$, ${{\mathbf{u}}_{1}}$, and ${{\mathbf{u}}_{2}}$ by the form
\begin{equation}
	\mathbf{u}={{\mathbf{u}}_{0}}-\left( {{\mathbf{u}}_{1}},{{\mathbf{u}}_{2}} \right){{\left( \mathbf{I}+\mathbf{R}_{2}^{T}\left( {{\mathbf{u}}_{1}},{{\mathbf{u}}_{2}} \right) \right)}^{-1}}\mathbf{R}_{2}^{T}{{\mathbf{u}}_{0}},
 \label{LINSYS_Sherman_Morrison2}
\end{equation}
where ${{\left( \mathbf{I}+\mathbf{R}_{2}^{T}\left( {{\mathbf{u}}_{1}},{{\mathbf{u}}_{2}} \right) \right)}^{-1}}\mathbf{R}_{2}^{T}{{\mathbf{u}}_{0}}$ can be computed cheaply because $\mathbf{R}_{2}^{T}\left( {{\mathbf{u}}_{1}},{{\mathbf{u}}_{2}} \right)$ is a $2\times 2$ matrix.

Based on the above derivation, the linear system (\ref{LINSYS_SCAL}) can be solved. The first two equations in linear system (\ref{LINSYS_SYM}), where the coefficient matrices are symmetric and typically sparse, are solved firstly. Then the results are combined with cheap operations. 

Another way to utilize the sparse structure is the sparsification of SOCs. The $i^{th}$ SOC $\mathcal{K}_{S}^{{{n}_{i}}}$ generates a ${{n}_{i}}\times {{n}_{i}}$ dense-matrix block in matrix $\mathbf{D}$ in terms of (\ref{DEF_G}) and (\ref{DEF_D}), which leads to heavy computation when dimension ${{n}_{i}}$ is large in direct methods for linear systems. High-dimension SOCs should be sparsified utilizing their internal sparsity to lessen computation.

The matrix block in $\mathbf{D}$ corresponding to  $\mathcal{K}_{S}^{{{n}_{i}}}$ is denoted by  $\mathbf{D}^{(i)}$. By definitions (\ref{DEF_G}) and (\ref{DEF_D}), and the property of Nesterov-Todd scalings (\ref{PROP_NT}),  $\mathbf{D}^{(i)}$ satisfies 
\begin{equation}
	\begin{aligned}
		&{{\left( {{\mathbf{D}}^{(i)}} \right)}^{2}}={{\left( {{\theta }^{(i)}}{{\mathbf{G}}^{(i)}} \right)}^{-2}}\\
		&={{\left( {{\theta }^{(i)}} \right)}^{-2}}\left( -{{\mathbf{Q}}^{(i)}}+2\left( {{\mathbf{Q}}^{(i)}}{{\mathbf{q}}^{(i)}} \right){{\left( {{\mathbf{Q}}^{(i)}}{{\mathbf{q}}^{(i)}} \right)}^{T}} \right)
	\end{aligned}
\end{equation}

Then, the linear system
\begin{equation}
	{{\left( {{\mathbf{D}}^{(i)}} \right)}^{2}}{{\mathbf{h}}_{1}}={{\mathbf{h}}_{2}}
\end{equation}
for vectors ${{\mathbf{h}}_{1}},{{\mathbf{h}}_{2}}\in {{R}^{{{n}_{i}}}}$ is equivalent to
\begin{equation}
	\begin{aligned}
		&{{\mathbf{h}}_{2}}={{\left( {{\theta }^{(i)}} \right)}^{-2}}\left( -{{\mathbf{Q}}^{(i)}}{{\mathbf{h}}_{1}}+2\left( {{\mathbf{p}}^{(i)}} \right){{\left( {{\mathbf{p}}^{(i)}} \right)}^{T}}{{\mathbf{h}}_{1}} \right)\\
		&={{\left( {{\theta }^{(i)}} \right)}^{-2}}\left( -{{\mathbf{Q}}^{(i)}}{{\mathbf{h}}_{1}}+\sqrt{2}{{h}_{3}}{{\mathbf{p}}^{(i)}} \right),
		\label{TRANS_SOC_CONSTRAIN}
	\end{aligned}
\end{equation}
where
\begin{equation}
	\begin{aligned}
		& {{\mathbf{p}}^{(i)}}={{\mathbf{Q}}^{(i)}}{{\mathbf{q}}^{(i)}} \\ 
		& {{h}_{3}}=\sqrt{2}{{\left( {{\mathbf{p}}^{(i)}} \right)}^{T}}{{\mathbf{h}}_{1}}.
	\end{aligned}
\end{equation}

Equation (\ref{TRANS_SOC_CONSTRAIN}) can be written in the matrix form
\begin{equation}
	{{\widehat{\mathbf{D}}}^{(i)}}\left( \begin{matrix}
		{{\mathbf{h}}_{1}}  \\
		{{h}_{3}}  \\
	\end{matrix} \right)=\left( \begin{matrix}
		{{\mathbf{h}}_{2}}  \\
		0  \\
	\end{matrix} \right),
\end{equation}
where
\begin{equation}
	{{\widehat{\mathbf{D}}}^{(i)}}\triangleq {{\left( {{\theta }^{(i)}} \right)}^{-2}}\left( \begin{matrix}
		-{{\mathbf{Q}}^{(i)}} & \sqrt{2}{{\mathbf{p}}^{(i)}}  \\
		\sqrt{2}{{\left( {{\mathbf{p}}^{(i)}} \right)}^{T}} & -1  \\
	\end{matrix} \right).
\end{equation}

Matrix ${{\widehat{\mathbf{D}}}^{(i)}}$ is sparse since ${{\mathbf{Q}}^{(i)}}$ is a diagonal matrix. However, its dimension is ${{n}_{i}}+1$ instead of ${{n}_{i}}$. The additional row and column neutralize the advantage in sparsity with low ${{n}_{i}}$. We observed that
\begin{equation}
	\begin{aligned}
		& \text{nnz}\left( {{\left( {{\mathbf{D}}^{(i)}} \right)}^{2}} \right)=n_{i}^{2} \\ 
		& \text{nnz}\left( {{\widehat{\mathbf{D}}}^{(i)}} \right)=3{{n}_{i}}+1 \\ 
		& \text{nnz}\left( {{\left( {{\mathbf{D}}^{(i)}} \right)}^{2}} \right)>\text{nnz}\left( {{\widehat{\mathbf{D}}}^{(i)}} \right) \text{ if } {n}_{i}\ge 4.
	\end{aligned}
\end{equation}
Therefore, dense matrix block ${{\left( {{\mathbf{D}}^{(i)}} \right)}^{2}}$ in the ${{\mathbf{D}}^{2}}$ term of $\widehat{\mathbf{B}}$ is replaced with sparse-matrix block ${{\widehat{\mathbf{D}}}^{(i)}}$ when ${{n}_{i}}\ge 4$. Coefficient matrix $\widehat{\mathbf{B}}$ is transformed into matrix $\mathbf{B}$ in the form
\begin{equation}
    \mathbf{B}=\left( \begin{matrix}
        0 & {{\mathbf{A}}^{T}} & {{{\mathbf{\hat{I}}}}^{T}}  \\
        \mathbf{A} & 0 & 0  \\
        {\mathbf{\hat{I}}} & 0 & \widehat{\mathbf{D}}  \\
    \end{matrix} \right).
    \label{DEF_B}
\end{equation}
In the definition,
\begin{equation}
    \begin{aligned}
        & \widehat{\mathbf{D}}\triangleq \text{blkdiag}\left( {{\widetilde{\mathbf{D}}}^{(1)}},\cdots ,{{\widetilde{\mathbf{D}}}^{(l+m)}} \right) \\ 
        & {{\widetilde{\mathbf{D}}}^{(i)}}\triangleq \left\{ \begin{aligned}
            & {{\widehat{\mathbf{D}}}^{(i)}},\text{      if }{{n}_{i}}\ge 4 \\ 
            & {{\left( {{\mathbf{D}}^{(i)}} \right)}^{2}},\text{ if }{{n}_{i}}<4, \\ 
        \end{aligned} \right. 
    \end{aligned}
    \label{DEF_D_HAT}
\end{equation}
where a 1D linear cone is viewed as a 1D SOC, and $\mathbf{\hat{I}}$ is identity matrix $\mathbf{I}$ with additional rows of zeros corresponding to additional rows in $\widehat{\mathbf{D}}$. The two lower right diagonal ones in $\widehat{\mathbf{B}}$ are removed since corresponding equations can be solved immediately.

The the first two equations in linear system (\ref{LINSYS_SYM}) are transformed into
\begin{equation}
	\begin{aligned}
		& \mathbf{B}{{{\mathbf{\hat{w}}}}_{1}}={{{\mathbf{\tilde{w}}}}_{1}} \\ 
		& \mathbf{B}{{{\mathbf{\hat{w}}}}_{2}}={{{\mathbf{\tilde{w}}}}_{2}},
	\end{aligned}
	\label{LINSYS_FINAL}
\end{equation}
where the right-hand sides ${{\mathbf{\tilde{w}}}_{1}}$ and ${{\mathbf{\tilde{w}}}_{2}}$ denotes ${{\mathbf{w}}_{0}}$ and ${{\left( {{\mathbf{R}}_{1}} \right)}_{1}}$ with additional elements of zeros corresponding to the additional rows in $\widehat{\mathbf{D}}$, respectively, and the last two terms are removed.

Solving linear system (\ref{LINSYS_FINAL}) is the most time-consuming in the IPM. Fortunately, the two equations share the same coefficient matrix, so the decomposition can be reused to reduce computation.

An efficient customized solver is developed to solve indefinite linear systems (\ref{LINSYS_FINAL}) by LDL decomposition with dynamic regularization and iterative refinement \cite{Domahidi2013}. The approximate minimum degree (AMD) method \cite{Amestoy2004} is applied to compute permutations, which reduces nnz after decomposition. Symbolic decomposition is performed to exploit the sparsity of given problem structures, which are reusable when the problem structures are unchanged. It is computed before real-time missions, and saved in files consisting of sequences of operand positions, which are loaded into memory before solving. The coefficient matrix is decomposed in the solving process according to the operand positions saved, so redundant operations to process zero elements are avoided. The manner is as efficient as the code generation approach \cite{Dueri2017, Mattingley2012} because the operations executed are roughly the same. However, the codes do not change with problem structures, so no recompiling is required to accommodate different problem sizes.

The method presented to solve linear system (\ref{LINSYS_SCAL}) is summarized as Algorithm \ref{ALG_SOLV_LINSYS}. It accelerates the solution of linear systems in lines 4 and 8 in Algorithm \ref{ALG_CLASSIC}, which consumes most of the computation in the IPM. APDG experiments in Section \uppercase\expandafter{\romannumeral5} shows that the method decreases coefficient matrix nnz significantly and computation sharply compared with the classic approach.

\begin{algorithm}[tb]
	\caption{Method to accelerate the solution of linear systems in the classic IPM for SOCP}
	\label{ALG_SOLV_LINSYS}
	\textbf{Input}: problem parameters $\mathbf{A}$, $\mathbf{b}$, $\mathbf{c}$, and $\mathcal{K}$, current estimation of solution $\mathbf{z}=(\mathbf{x},\mathbf{y},\mathbf{s},\kappa,\tau)$, additional terms $\mathbf{E}_{xs}, {E}_{\kappa \tau }$, and $\nu$.\\
	\textbf{Output}: Vector $\mathbf{u}=(\Delta\mathbf{x},\Delta\mathbf{y},\Delta\mathbf{s},\Delta\kappa,\Delta\tau)^T$.
	\begin{algorithmic}[1] 
            \STATE Calculate $\mathbf{\hat{D}}$ by (\ref{DEF_D_HAT}) and record the additional row numbers as $\mathcal{I}_a$;\\
            \STATE Set the coefficient matrix $\mathbf{B}$ by (\ref{DEF_B});\\
            \STATE Calculate $\mathbf{w}_0$ by (\ref{LINSYS_MAT1}) and $(\mathbf{R}_1)_1$ by (\ref{LINSYS_MAT1_REFORM});\\
            \STATE Set $\mathbf{\tilde{w}}_1$, $\mathbf{\tilde{w}}_2$ by adding zeros in terms $\mathcal{I}_a$ of $\mathbf{w}_0$, $(\mathbf{R}_1)_1$, and remove the last 2 terms, respectively;\\
            \STATE Solve linear system (\ref{LINSYS_FINAL}) to obtain $\mathbf{\hat{w}}_1$ and $\mathbf{\hat{w}}_2$;\\
            \STATE Set $\mathbf{u}_0$ and $\mathbf{u}_1$ by removing terms $\mathcal{I}_a$ of $\mathbf{\hat{w}}_1$ and $\mathbf{\hat{w}}_2$, respectively;
            \STATE Append the last 2 terms of $\mathbf{w}_0$ and $(\mathbf{R}_1)_1$ to $\mathbf{u}_0$ and $\mathbf{u}_1$, respectively;\\
            \STATE Calculate $\mathbf{u}$ by (\ref{LINSYS_Sherman_Morrison2}).
	\end{algorithmic}
\end{algorithm}

\section{Warm-starting for Correlated SOCP Subproblems}

This section presents a warm-starting scheme that significantly accelerates the solution of correlated SOCP problems in scenarios such as successive convexification. The scheme uses rough estimates of the previous solution to generate warm-starting points, which enables each subproblem to be processed for only a few iterations. 

It is widely perceived that warm-starting of IPMs is difficult \cite{Potra2000}. If the solution to the previous problem is on the boundary of the feasible region, it may be also close to the boundary in the new problem. When the estimate of the solution is close to the boundary, the gradients and Hessians of the barrier functions change rapidly. Therefore, IPMs generally behave poorly if the solution is not well-centered, which produces either ill-conditioned linear systems or noneffective searching directions \cite{Skajaa2013}. Solutions are typically on the boundary in many real problems, so cold-starting usually performs better than using the previous solution directly for warm-starting.

Ref. \cite{Skajaa2013} presents a warm-starting scheme by initializing with a linear combination of the optimal solution of a previous problem and the cold-starting point with a predefined weight, which keeps the initial value away from the boundary of the feasible region. The scheme is extended by using an inexact solution to the previous problem and a problem-dependent weight.

The previous SOCP is denoted by ${{\mathcal{P}}^{o}}$, and its parameters are denoted by ${{\mathbf{A}}_{o}},{{\mathbf{b}}_{o}},{{\mathbf{c}}_{o}}$. The current SOCP is $\mathcal{P}$ and its parameters are $\mathbf{A},\mathbf{b},\mathbf{c}$. The two problems share the same cone constraint $\mathbf{K}$. $\left( {{\mathbf{x}}_{o}},{{\mathbf{y}}_{o}},{{\mathbf{s}}_{o}} \right)$ is an inexact solution of ${{\mathcal{P}}^{o}}$. Then, warm-starting point ${{\mathbf{z}}_{w}}=\left( {{\mathbf{x}}_{w}},{{\mathbf{y}}_{w}},{{\mathbf{s}}_{w}},{{\kappa }_{w}},{{\tau }_{w}} \right)$ is calculated as
\begin{equation}
	\begin{aligned}
		& {{\mathbf{x}}_{w}}=\lambda {{\mathbf{x}}_{o}}+\left( 1-\lambda  \right)\mathbf{e} \\ 
		& {{\mathbf{y}}_{w}}=\lambda {{\mathbf{y}}_{o}} \\ 
		& {{\mathbf{s}}_{w}}=\lambda {{\mathbf{s}}_{o}}+\left( 1-\lambda  \right)\mathbf{e} \\ 
		& {{\kappa }_{w}}={{\left( {{\mathbf{x}}_{o}} \right)}^{T}}{{\mathbf{s}}_{o}}/k \\ 
		& {{\tau }_{w}}=1,
	\end{aligned}
\end{equation}
where the weight
\begin{equation}
	\lambda =\max \left( 1-1/\left( {{\left\| \mathbf{A} \right\|}_{\infty }}+{{\left\| \mathbf{b} \right\|}_{\infty }} \right),{{\lambda }_{0}} \right).
\end{equation}
$\lambda_0$ is a predefined parameter (0.999 by default).

The work studies the conditions under which the warm-starting scheme improves the worst-case iteration complexity and summarize the results in Ref. \cite{Chen2022}. The research serves as the theoretical basis of the scheme, but it is too long to be included in this work. We prove that an infeasible IPM for SOCPs compatible with the warm-starting scheme, has $O\left(\sqrt{k}\log\left(1/\epsilon\right)\right)$ worst-case iteration complexity to obtain a solution or an infeasibility certificate. Although the complexity is the same as the best-known worst-case complexity \cite{Monteiro2000} of IPMs for SOCPs, it had only been proven for several feasible IPMs that are inconvenient for warm-starting. When weight $\lambda$ is close to 1, the warm-starting scheme can reduce required iterations compared with cold-starting.

\section{Simulation Results}

The section presents numerical results to demonstrate the effectiveness and performance of the IPMs proposed to solve APDG. Firstly, a sample scenario of powered rocket landing is introduced to investigate the correctness of the solution and the computational efficiency. Secondly, Monte Carlo simulations are performed to evaluate the performance of different IPM solvers. The experiments are run on a workstation with an AMD Ryzen 7 5800H CPU (3.2-4.4GHz).

\subsection{Sample Scenario of Powered Rocket Landing}

A sample scenario of powered rocket landing is presented. The following assumptions are made: 1) Force acting on the vehicle are thrust, gravity, and aerodynamic drag, and lift is negligible. 2) The vehicle is sufficiently close to the surface, so surface curvature and changes in gravity are ignored. 3) The bandwidth of vehicle’s attitude control is sufficiently high to decouple the translational and rotational dynamics \cite{Dueri2017}. Therefore, the vehicle is modeled as a 3 degree-of-freedom (DOF) point-mass subject under the last assumption and does not include attitude dynamics. Inequity constraints include the max velocity, fuel mass limits, glide-slope cone, commanded thrust range, maximum throttling rate, and maximum tilt angle, see \cite{Szmuk2016} for details.

A Cartesian coordinate system is used, and its origin is the landing point. X ,Y, and Z directions are east, north, and up, respectively. Variables are defined as follows: $\mathbf{r}\in {{R}^{3}}$ is the position; $\mathbf{v}\in {{R}^{3}}$ is the velocity; $\mathbf{a}\in {{R}^{3}}$ is acceleration; $\mathbf{T}\in {{R}^{3}}$ is thrust; ${{\mathbf{D}}_{a}}\in {{R}^{3}}$ is aerodynamic drag; $\mathbf{g}\in {{R}^{3}}$ is gravity acceleration; $m$ is mass; $t$ is time; ${{t}_{f}}$ is the final time; $\rho $ is the air density; ${{I}_{sp}}$ is the specified impulse of the rocket motor; ${{S}_{ref}}$ is the drag reference area; ${{C}_{D}}$ is the coefficient of drag; ${{\rho }_{0}}$ is the air density at the landing position; ${{m}_{dry}}$ is dry mass; ${{v}_{\max }}$ is the maximum speed; $\left[ {{T}_{\min }},{{T}_{\max }} \right]$ is the range of thrust magnitude; $\left[ {{{\dot{T}}}_{\min }},{{{\dot{T}}}_{\max }} \right]$ is the range of the thrust changing rate; ${{\theta }_{T,max}}$ is the maximum tilt angle; ${{\theta }_{gs}}$ is the maximum gliding-slope cone angle; ${{\mathbf{r}}_{\mathbf{0}}},{{\mathbf{v}}_{0}}$ ${{m}_{0}}$, and ${{t}_{f,0}}$ are the initial values of $\mathbf{r},\mathbf{v},m$, and ${{t}_{f}}$.

The aerodynamic drag is expressed as
\begin{equation}
    {{\mathbf{D}}_{a}}(t)=-\frac{{{C}_{D}}{{S}_{ref}}}{2}{{\rho }_{0}}\exp \left( -{{c}_{rho}}{{\mathbf{r}}_{y}} \right)\left\| \mathbf{v}(t) \right\|\mathbf{v}(t),
\end{equation}							
where the air density decays exponentially according to the altitude ${{\mathbf{r}}_{y}}$, and ${{c}_{rho}}$ is a constant.

The fuel consumption dynamic is given by 
\begin{equation}
    \dot{m}=-\left\| \mathbf{T}(t) \right\|/\left( {{I}_{sp}}\mathbf{g} \right).
\end{equation}
						
The problem is transformed into a sequence of SOCPs using the successive convexification approach \cite{Szmuk2016} with minor modifications as follows: 1) When upper bounds of the thrust regions and acceleration error are introduced as penalty terms in the objective function, the average value on each time grid point is used instead of the root square sum, which avoids the application of high-dimension SOC constraints to save computation. 2) The trajectory obtained in each successive convexification step (hereinafter referred to as the SC step) is verified by numerical simulations using the programmed thrust profile on fine time grids. The algorithm terminates when the simulated landing position and velocity errors are below predefined error bounds. The parameters required are defined as follows: ${{\omega }_{m,f}}$, ${{\omega }_{\eta ,\Delta t}}$, ${{\omega }_{\eta ,T}}$, and ${{\omega }_{\kappa ,a,R}}$ are the coefficients in successive convexification, corresponding to mass, change of time step, change of trust, and acceleration error, respectively. ${{k}_{f}}$ and ${{k}_{fine}}$ are the number of time steps in the original and fine grids, respectively; $\epsilon_r$ and $\epsilon_v$ are the error bounds of landing positions and velocity errors, respectively.

\begin{table*}
\caption{Parameters of the Powered Rocket Landing Experiment}
\label{TAB_PARAM}
\tablefont
\centering
\begin{tabular}{llllllll}
\toprule
Parameter      &Value       &Parameter      &Value          &Parameter      &Value          &Parameter      &Value\\ 
\midrule
$\rho_0$   &1.225 kg/m$^3$   &$c_{rho}$    &0.0001      &$S_{ref}$      &10 m$^2$    &$\mathbf{g}$ &[0,~-9.8,~0]$^T$ m/s$^2$ \\
$C_D$   &0.5      &$m_{dry}$   &30,000 kg       &$I_{sp}$       &300 s        &$\mathbf{r}_0$ &[-1,000,~4,000,~500]$^T$ m\\
$v_{max}$ &340 m/s  &$m_0$  &40,000 kg    &$t_{f,0}$  &35 s     &$\mathbf{v}_0$ &[-50,~-200,~-100]$^T$ m/s\\
$T_{min}$   &300 kN   &$T_{max}$ &1,000 kN  &$\dot{T}_{min}$    &-100 kN/s    &$\dot{T}_{max}$    &100 kN/s\\
$\theta_{T,max}$    &30$^\circ$ &$\theta_{gs}$    &80$^\circ$   &$N_{iter}$ &60   &$k_f$  &30\\
$\omega_{m,f}$  &1.0 kg$^{-1}$   &$\omega_{\eta,\Delta t}$  &0.1 s$^{-1}$ &$\omega_{\eta,T}$  &0.01 kN$^{-1}$ &$\omega_{\kappa,a,R}$  &500,000 s$^2$/m\\
$k_{fine}$  &300  &$\epsilon_r$   &2 m      &$\epsilon_v$ &0.2 m/s\\
\bottomrule
\end{tabular}
\end{table*}

\begin{figure}
\centerline{\includegraphics[width=21pc]{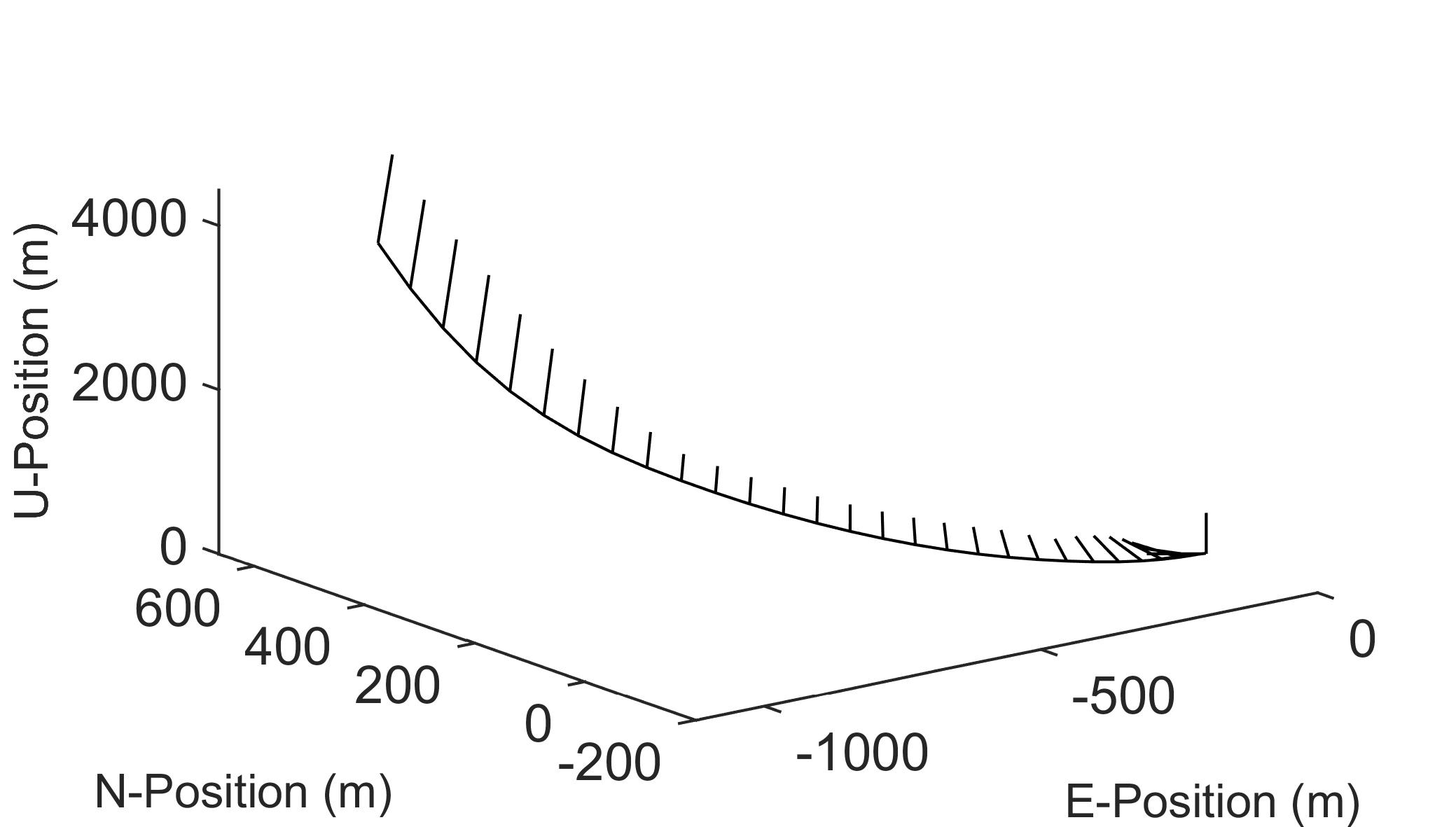}}
\caption{3D overview of the fuel-optimal trajectory. Lines intersecting the trajectory represent scaled thrust vectors.}
\label{FIG_TRAJ}
\end{figure}

\begin{figure*}
\centerline{\includegraphics[width=43pc]{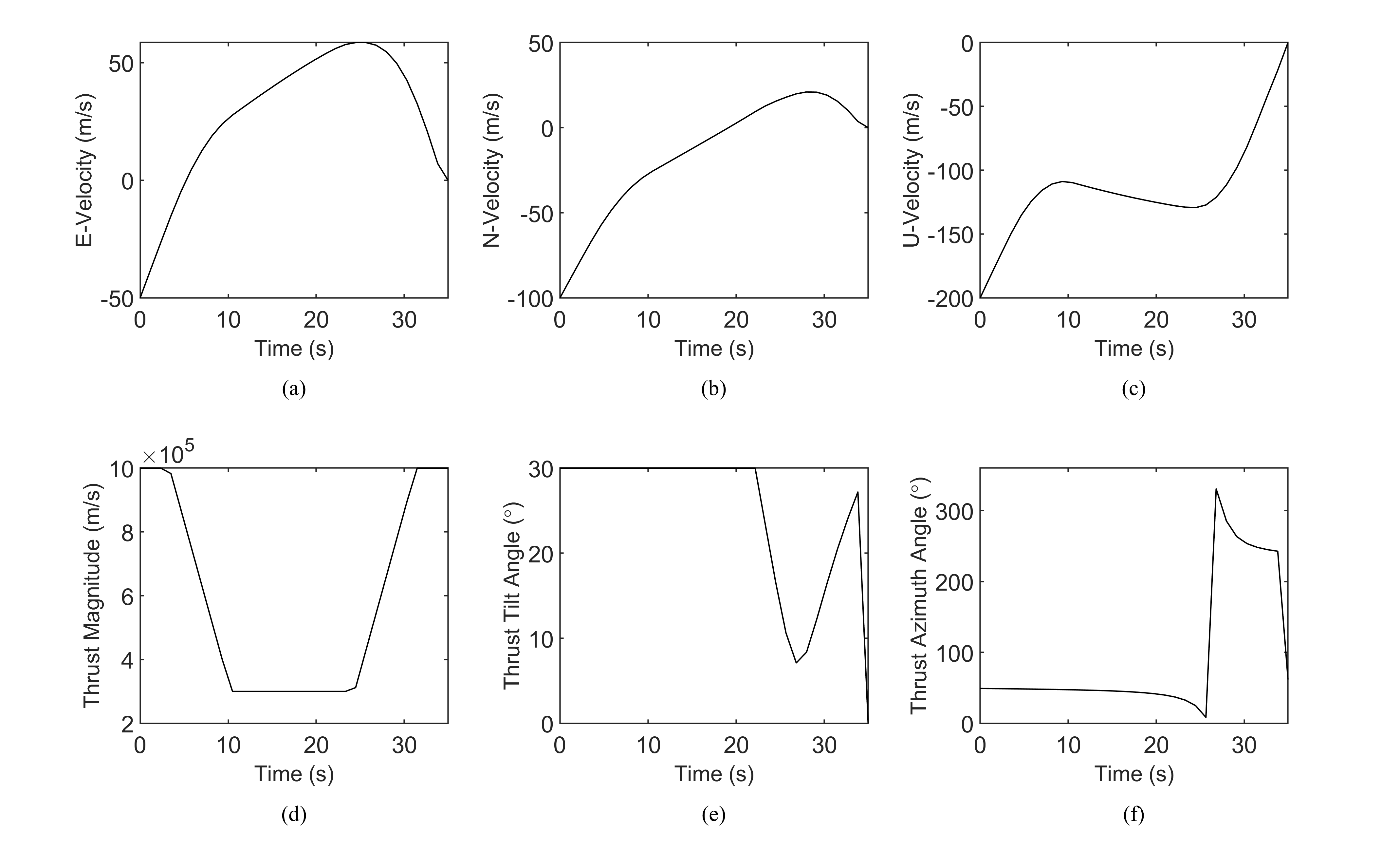}}
\caption{Velocities and thrust: (a), (b), and (c) are the east, north, and up components of the velocities, respectively; (d), (e), and (f) are the magnitude, tilt angle, and azimuth angle of the thrusts.}
\label{FIG_V_T}
\end{figure*}

\begin{figure}
\centerline{\includegraphics[width=21pc]{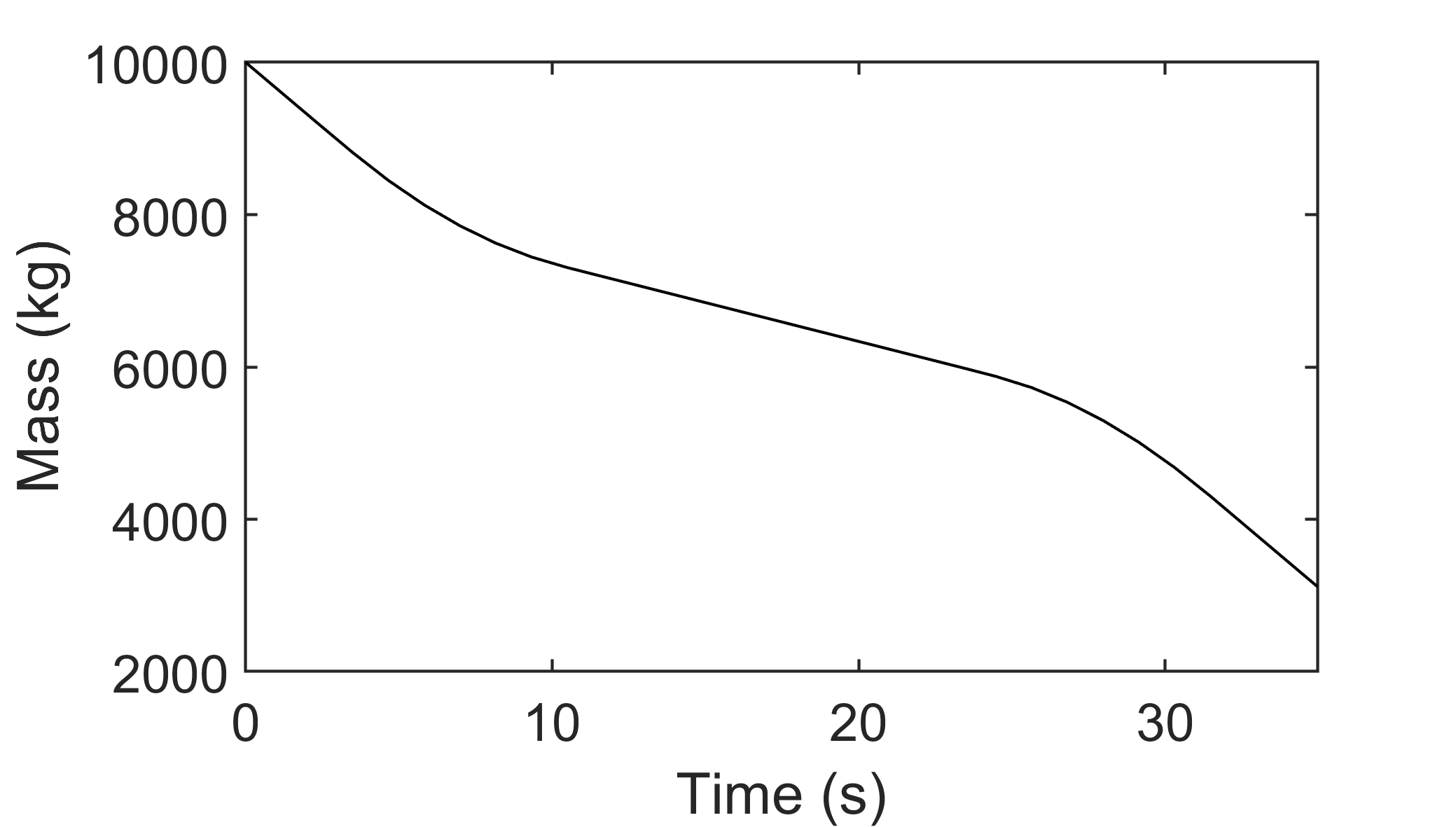}}
\caption{Mass time history.}
\label{FIG_M}
\end{figure}

The parameters of the experiment are listed in Table. \ref{TAB_PARAM}. Figs. \ref{FIG_TRAJ}, \ref{FIG_V_T}, and \ref{FIG_M} present the programmed trajectory, velocity, trust, and mass obtained by our algorithm (FSOCP). The results are obtained with a fast version of warm-starting (called 1-step warm starting), in which each SOCP problem is solved with only 1 iteration step. There are a total of 40 steps in the example. The trajectory ends at the origin point, with zero speed and a vertical thrust vector, showing that vertical soft-landing is performed (see Fig. \ref{FIG_TRAJ}). Thrust performs bang-coast-bang maneuver (see Fig. \ref{FIG_V_T} (d)). Fig. \ref{FIG_V_T} (e) shows that the thrusts tilt angles are bounded by $\theta_{T,max}=30^\circ$, although they seem to be large in Fig. \ref{FIG_TRAJ} due to small vertical scale. The maximum tilt angle constraint is activated during the first stage of descending. 

Then, the computation time of FSOCP is compared with the results of SDPT3 \cite{Toh1999}, MOSEK \cite{Andersen2000}, ECOS \cite{Domahidi2013}, and Coneprog \cite{Coneprog}. The experiment runs 100 times, and the average results are presented. FSOCP is implemented in C, and other solvers are called in Matlab. Since Matlab scripts are typically slower than C, we include all the run time for our algorithm and consider only the time to solve SOCPs for other solvers to make fair comparisons. Then, the reported run time of MOSEK and ECOS are unaffected by Matlab, because they solve SOCPs solely with mex files, which are efficient libraries written in C, C++, or Fortran. We also ignore the setup time for ECOS, which is reusable according to its algorithm \cite{Domahidi2013}. Part of the work for SDPT3 and Coneprog still needs to be completed with Matlab scripts, which adversely affects their computational efficiency.

\begin{table*}
\caption{Comparison of Solvers in the Powered Rocket Landing Experiment}
\label{TAB_COMP_SOLVER}
\tablefont
\centering
\begin{tabular}{llllll}
\toprule
Solver	&Run time (ms)	&SC step	&Position error (m)	&Velocity error (m/s)	&Fuel remained (kg)\\
\midrule
FSOCP (c)	&27.0	&\textbf{3}	&0.672	&\textbf{0.068}	&\textbf{3,123.9}\\
FSOCP (w1)	&\textbf{10.0}	&40	&0.684	&0.069	&3,123.2\\
SDPT3	&3,177.1	&4	&0.685	&0.071	&3,114.8\\
MOSEK	&68.2	&4	&0.677	&0.069	&3,114.9\\
ECOS	&158.5	&6	&1.893	&0.134	&2,247.9\\
Coneprog	&8,419.2	&4	&0.937	&0.104	&3,078.0\\
FBSOCP	&456.9	&\textbf{3}	&\textbf{0.664}	&\textbf{0.068}	&3,123.7\\
\bottomrule
\end{tabular}
\end{table*}

Table. \ref{TAB_COMP_SOLVER} shows the run time. FSOCP (c) and (wn) denote FSOCP with cold-starting and n-step warm-starting, respectively. FSOCP (c) is more than twice faster than MOSEK in the experiment, which is the fastest among the publicly available solvers. FSOCP (w1) further accelerates roughly by a factor of 3, showing that warm-starting is effective in successive convexification. It requires more SC steps, but consumes low run time, because only 1 iteration is performed in each SC step. MOSEK and ECOS are also efficient in computation, and the former is faster. SDPT3 and Coneprog are much slower than other solvers. Table. \ref{TAB_COMP_SOLVER} also shows the fuel cost and landing error obtained by numerical simulations on fine grids. Landing errors of different solvers do not vary a lot, because the successive convexification terminates when the errors meet the requirements. The fuels remaining for FSOCP, SDPT3, MOSEK, Coneprog, and FBSOCP are roughly the same, which exhibits fuel optimality, but ECOS requires much more fuels.  

The work develops FBSOCP, a variant of FSOCP using the classic IPM \cite{Andersen2003, Wang2003, Dueri2017}, to verify the effectiveness of our approach for accelerating the solution of linear systems. Another example is added by ignoring the aerodynamic forces and keeping other parameters unchanged (hereinafter called NAPDG), and the problem is converted to a single SOCP according to Ref. \cite{Dueri2017}. The coefficient matrix sparsity of FSOCP and FBSOCP in the two experiments are compared in Table. \ref{TAB_COMP_SPARSITY}. The run time of FSOCP, MOSEK, ECOS, and FBSOCP in NAPDG is 2.55, 11.1, 19.3, and 4.86 ms, respectively. The classic IPM is efficient in NAPDG, which matches the results in Ref. \cite{Dueri2017}. FSOCP is faster than FBSOCP, because it reduces the number of nonzero elements (nnz) of the coefficient matrix, and increases the dimension, which makes the problem sparser. The efficiency of FBSOCP severely degrades since nnz surges 9.7 times in APDG. FSOCP (c) is 16.9 times faster than FBSOCP because it improves sparsity by reducing nnz and increasing the dimension significantly.

\begin{table*}
\caption{Comparision of sparsity of FSOCP and the classic IPM (FBSOCP)}
\tablefont
\label{TAB_COMP_SPARSITY}
\centering
\begin{tabular}{lllllll}
\toprule
        &\multicolumn{3}{c}{PDG without aerodynamic forces}          &\multicolumn{3}{c}{PDG with aerodynamic forces}\\
\cmidrule(r){2-7}
Solver	&nnz	            &Dimension          &Run time (ms)      &nnz                &Dimension	     &Run time (ms)\\
\midrule
FSOCP (c)   &\textbf{5,567}     &\textbf{1,478}     &\textbf{2.55}      &\textbf{12,102}    &\textbf{2,397} &\textbf{27.0}\\
FBSOCP  &8,455              &437                &4.86               &82,201             &533            &456.9\\
\bottomrule
\end{tabular}
\end{table*}

\begin{figure}
\centerline{\includegraphics[width=21pc]{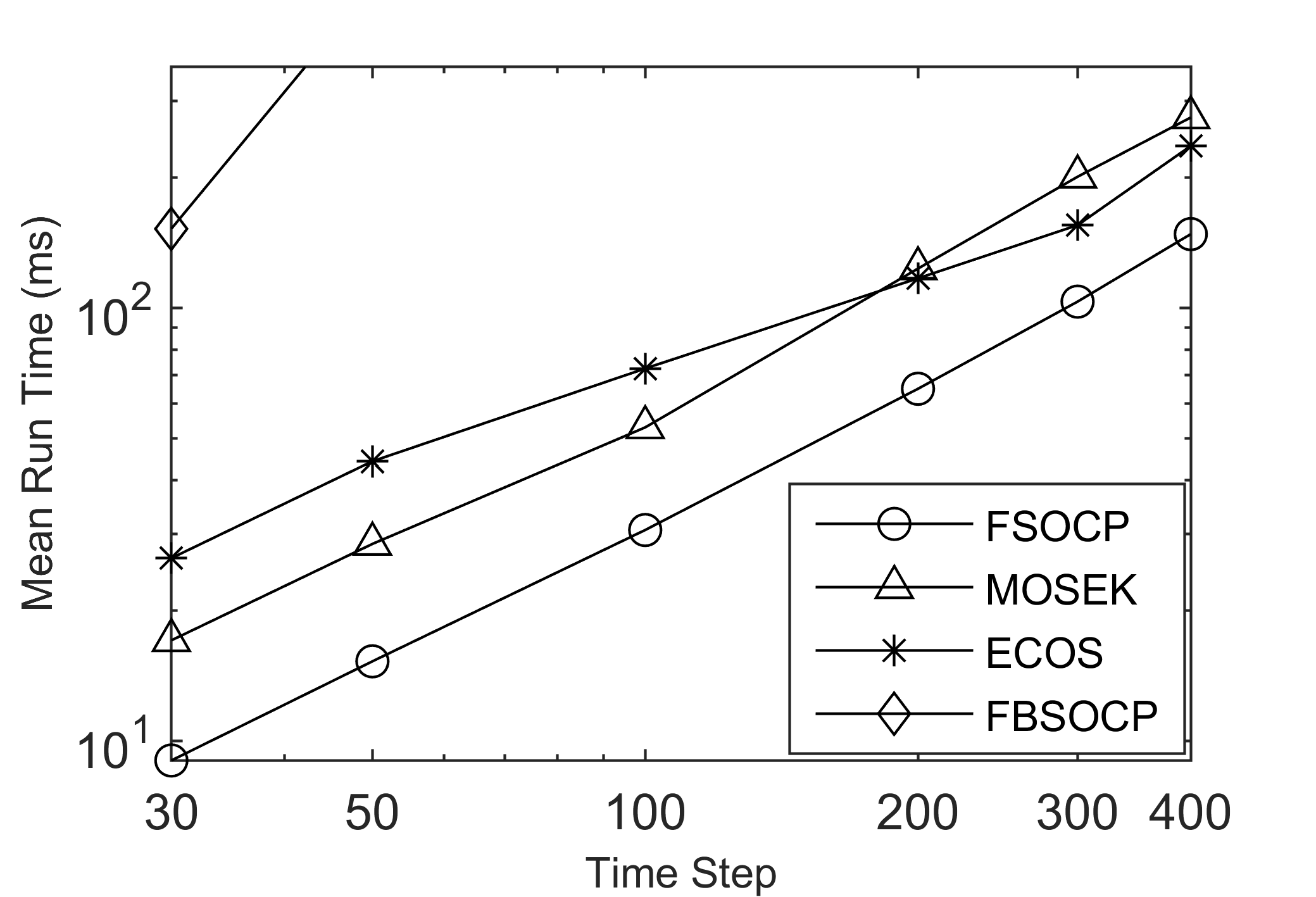}}
\caption{Mean run time per SOCP for different problem sizes.}
\label{FIG_SCAL}
\end{figure}

Fig. \ref{FIG_SCAL} compares the running times per SOCP of solvers for different problem sizes, and Table. \ref{TAB_COMP_SCAL} presents the corresponding configuration of SOCPs. FSOCP (cold-starting), MOSEK, and ECOS scale well as time steps increase. FSOCP is the fastest for all configurations. MOSEK is more efficient than ECOS in small problems, but the latter scales better. Coneprog and SDPT3 are not included in Fig. \ref{FIG_SCAL} because their run time is too long.

\begin{table}
\caption{Configuration of SOCPs for different problem sizes}
\label{TAB_COMP_SCAL}
\tablefont
\centering
\begin{tabular}{llll}
\toprule
Time steps	&Solution variables	 &LCs	&SOCs\\
\midrule
30	&870	&281	&155\\
50	&1,430	&461	&255\\
100	&2,830	&911	&505\\
200	&5,630	&1,811	&1,005\\
300	&8,430	&2,711	&1,505\\
400	&11,230	&3,611	&2,005\\
\bottomrule
\end{tabular}
\end{table}

\subsection{Monte Carlo Simulation}

The performance of solvers is evaluated by Monte Carlo simulations with random initial conditions obtained by adding zero-mean Gaussian noise to the initial condition in Table 1. The standard deviation of noises added on the initial position components, velocity components, and fuel mass are 500 m, 50 m/s, and 300 kg, respectively. Other parameters in the powered rocket landing experiment are unchanged. The maximum number of SC steps is 120 for 1-step warm-starting of FSOCP, and 30 for other cases. The algorithm terminates and returns with a failure when the maximum SC step number is exceeded.

The experiment runs 10,000 times and the average results are reported in Table. \ref{TAB_COMP_SOLVER_MC}. The success rate and fuel remaining of FSOCP and MOSEK are roughly at the same level, which is much larger than those of ECOS. FSOCP (c) is faster and generates a higher success rate compared with MOSEK. FSOCP (w1) has the highest computational efficiency, and its run time is 10.5\% of MOSEK and 3.4\% of ECOS. Although its success rate is slightly lower than that of MOSEK, the latter has to predict the initial value in real cases for a much larger time offset (e.g., several seconds on a flight processor) in the presence of random or unpredictable factors, which may generate an unacceptable error. FSOCP (w5) is still 7 times faster than MOSEK, and the gap in the success rate is much smaller.

\begin{table}
\caption{Comparison of solvers in the Monte Carlo simulation}
\tablefont
\label{TAB_COMP_SOLVER_MC}
\centering
\begin{tabular}{lllll}
\toprule
Solver	&Success	&Run time	&SC	      &Fuel remained\\
        &rate	    &(ms)	    &step	  &(kg)\\
\midrule
FSOCP (c)	&\textbf{84.2}\%	&65.7	&7.4	&2,953.8\\
FSOCP (w1)	&81.8\%	&\textbf{8.2}	&32.3	&2,984.7\\
FSOCP (w5)	&82.5\%	&10.3	&9.7	&\textbf{2,998.0}\\
MOSEK	&83.3\%	&78.0	&\textbf{4.84}	&2,985.6\\
ECOS	&59.1\%	&240.9	&9.68	&2,416.8\\
\bottomrule
\end{tabular}
\end{table}

The average run time of FSOCP (w1) is approximately 0.6 s on a P2020NXE2KHC radiation-hardened flight processor (1 GHz), where APDG takes 1 of the 2 CPU cores. The result shows the algorithm is suitable for onboard implementation \cite{Scharf2015}. It is worth noting that simply scaling the 8.2 ms runtime from Table. \ref{TAB_COMP_SOLVER_MC} by 4.4 GHz/ 1 GHz obtains 0.036 s, which underestimates the runtimes by an order of magnitude. The result is consistent with the findings in Ref. \cite{Dueri2017}. 

In contrast, the average run time of FBSOCP is approximately 30 s on P2020NXE2KHC. An existing fast solver may require several seconds onboard. Random or unpredictable factors may cause unacceptable errors during the long time offset, which may lead to the failure of landing missions. However, new control instructions are calculated based on the predicted initial state, making it difficult to compensate for unpredictable errors generated in the running time. The methods proposed reduce the time to update control instructions significantly. As a result, it may reduce the effect of unpredictable errors and ultimately enhance the success rate of landing missions.

\section{Conclusions}

The work presented a fast interior-point method for solving the SOCP subproblems in fuel optimal atomsperic powered descent guidance (APDG). The innovation points were twofold: 1) The solution of linear systems, which costs most of the computation, was accelerated by an algorithm to exploit the sparsity of the problem structure. 2) A warm-starting scheme was proposed to utilize the correlation between subproblems, which enabled each subproblem to be solved for a few iterations. The method proposed was efficient for the correlated convex subproblems obtained in successive convexification. It was 9 times faster than MOSEK in Monte Carlo simulations performed to evaluate the efficiency of solvers in APDG, while the latter was the fastest publicly available solver tested in the problem. It cost approximately 0.6 s on a radiation-hardened flight processor, demonstrating that the method is applicable to solve real-time onboard APDG. Additionally, the approach may be useful to accelerate the solution of correlated SOCPs in successive convexification for various applications.

\bibliography{myref}

\begin{IEEEbiography}[{\includegraphics[width=1in,height=1.25in,clip,keepaspectratio]{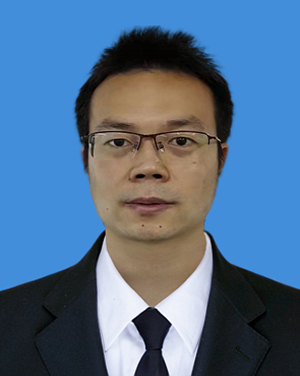}}]{Yushu Chen}{\space}received the M.S. degree in compute science from the National University of Defense Technology, Changsha, China, in 2007, and the Ph.D. degree in compute science from Tsinghua University, Beijing, China, in 2015. 

He is currently an Engineer with the Department of Computer Science and Technology, Tsinghua University. He also works at National Supercomputing Center in Wuxi, China. His research interests include numerical optimization, machine learning, numerical modeling, and guidance and control.
\end{IEEEbiography}%

\begin{IEEEbiography}[{\includegraphics[width=1in,height=1.25in,clip,keepaspectratio]{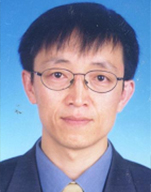}}]{Guangwen Yang}{\space}received the M.S. and Ph.D. degrees in applied mathmatics from the Harbin Institute of Technology, Harbin, China, in 1987 and 1996, respectively.

He is currently a Professor with the Department of Computer Science and Technology, Tsinghua University. He is also the Director of National Supercomputing Center in Wuxi, China. He also works at Zhejiang Lab, Hangzhou, China. His research interests include computer architecture, high performance computing, heterogeneous computing, and numerical algorithms.
\end{IEEEbiography}

\begin{IEEEbiography}[{\includegraphics[width=1in,height=1.25in,clip,keepaspectratio]{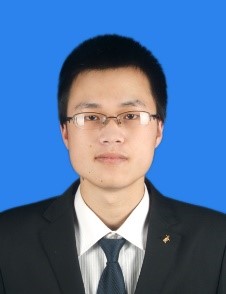}}]{Lu Wang}{\space}received the M.S. degree in aerospace engineering from the Northwestern Polytechnical University, Xi'an, China in 2019.

He is currently an Engineer with the Shanghai Aerospace Control Technology Institute. His research interests include convex optimization, guidance and control.
\end{IEEEbiography}

\begin{IEEEbiography}[{\includegraphics[width=1in,height=1.25in,clip,keepaspectratio]{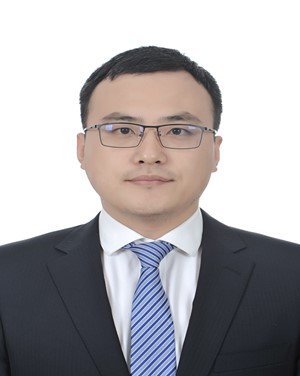}}]{Haipeng Chen}{\space}received the M.S. degree in navigation, guidance, and control from Harbin Institute of Technology, Harbin, China, in 2012. 
	
He is currently a senior engineer with Shanghai Aerospace Control Technology Institute, Shanghai, China. His main research interest lies in launch vehicle GNC system design.
\end{IEEEbiography}

\begin{IEEEbiography}[{\includegraphics[width=1in,height=1.25in,clip,keepaspectratio]{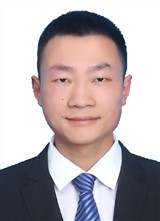}}]{Qingzhong Gan}{\space}received the B.S. degree and the M.S. degree in spacecraft design and engineering from the Harbin Institute of Technology, Harbin, China, in 2014 and 2017, respectively.
	
He is currently a Deputy chief designer of GNC system for reusable rockets with Shanghai Aerospace Control Technology Institute. He is involved in the development of systems, operations, and guidance algorithms for precision propulsive landing of booster stages, including the vertical takeoff and vertical landing experimental vehicle.
\end{IEEEbiography}

\begin{IEEEbiography}[{\includegraphics[width=1in,height=1.25in,clip,keepaspectratio]{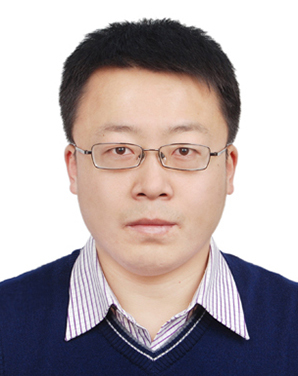}}]{Quanyong Xu}{\space}received the M.S. degree in aeronautical science and technology from the Beihang University, Beijing, China, in 2006, and the Ph.D. degree in aeronautical science and technology from Beihang University, Beijing, China, in 2010. 

He is currently an associate professor with the Institue for Aero Engine, Tsinghua University. His research interests include numerical modeling and simulation, aerodynamics, jet propulsion, and scientific computation. 
\end{IEEEbiography}

\end{document}